\documentclass[aps,prl,reprint,groupedaddress]{revtex4-1}
\usepackage{braket}
\usepackage[]{graphicx}
\usepackage{subfigure}
\usepackage[table]{xcolor}
\usepackage{amsbsy}
\usepackage{amsfonts}
\usepackage{MnSymbol}

\begin{document}

\title{4 $\times$ 20 Gbit/s mode division multiplexing over free space using vector modes and a $q$-plate mode (de)multiplexer}

\author{Giovanni Milione,$^{1,2,3,4*}$}
\author{Martin P. J. Lavery$^5$}
\author{Hao Huang$^6$}
\author{Yongxiong Ren$^6$}
\author{Guodong Xie$^6$}
\author{Thien An Nguyen$^{1,2}$}
\author{Ebrahim Karimi$^{7}$}
\author{Lorenzo Marrucci$^{4,8}$}
\author{Daniel A. Nolan$^{4,9}$}
\author{Robert R. Alfano$^{1,2,3,4}$} 
\author{Alan E. Willner$^6$} 

\affiliation{$^1$Institute for Ultrafast Spectroscopy and Lasers, New York, NY 10031 USA}
\affiliation{$^2$Physics Department, City College of the City University of  New York, New York, NY 10031 USA}
\affiliation{$^3$Physics Department, Graduate Center of the City University of New York, New York, NY 10016 USA}
\affiliation{$^4$New York State Center for Complex Light, New York, NY 10031 USA}
\affiliation{$^5$School of Engineering, University of Glasgow, Glasgow G12 8QQ, Scotland, UK}
\affiliation{$^6$Department of Electrical Engineering, University of Southern California, Los Angeles, CA 90089 USA}
\affiliation{$^7$Department of Physics, University of Ottawa, Ottawa, Ontario, K1N 6N5 Canada} 
\affiliation{$^8$Dipartimento di Fisica, Universita di Napoli Federico II, Complesso Universitario di Monte Sant'Angelo, Napoli, Italy}
\affiliation{$^9$Corning Incorporated, Sullivan Park, Corning, 14831 NY USA}

\affiliation{$^*$Corresponding author: giomilione@gmail.com}

\date{\today}

\begin{abstract}
Vector modes are spatial modes that have spatially inhomogeneous states of polarization, such as, radial and azimuthal polarization. They can produce smaller spot sizes and stronger longitudinal polarization components upon focusing. As a result, they are used for many applications, including optical trapping and nanoscale imaging. In this work, vector modes are used to increase the information capacity of free space optical communication via the method of optical communication referred to as mode division multiplexing. A mode (de)multiplexer for vector modes based on a liquid crystal technology referred to as a $q$-plate is introduced. As a proof of principle, using the mode (de)multiplexer four vector modes each carrying a 20 Gbit/s quadrature phase shift keying signal on a single wavelength channel ($\lambda \sim $1550nm), comprising an aggregate 80 Gbit/s, were transmitted $\sim$1m over the lab table with $<$-16.4 dB ($\sim 2 \%$) mode crosstalk. Bit error rates for all vector modes were measured at the forward error correction threshold with power penalties $<$ 3.41dB.
 \end{abstract}

\maketitle

\section{Introduction}
\begin{figure*}[htb]
\includegraphics[width=\linewidth]{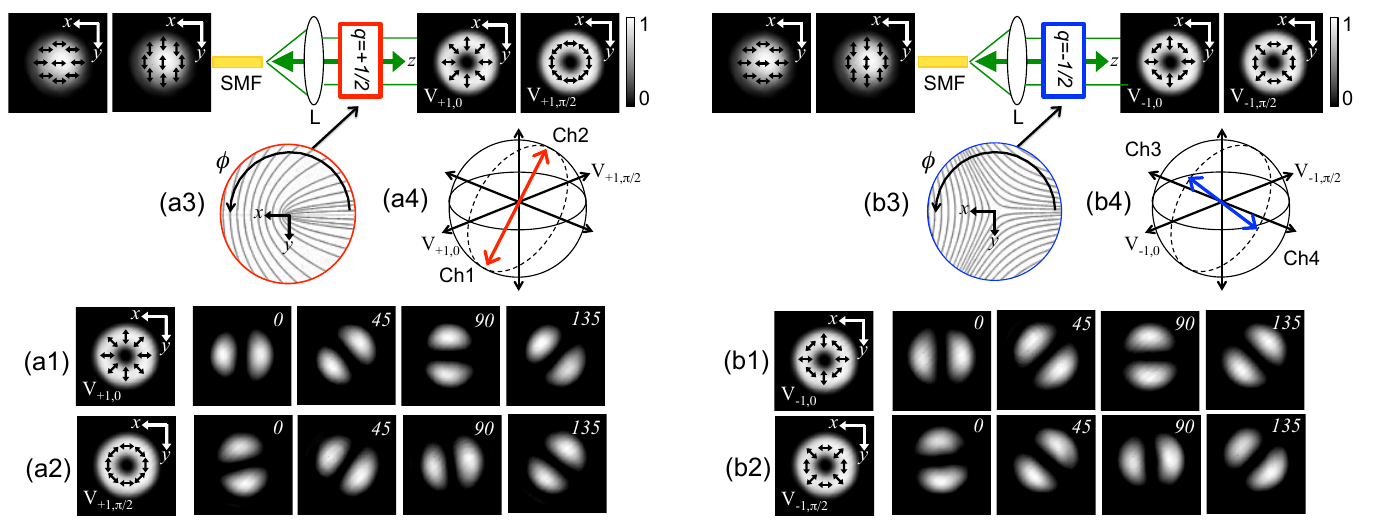}
\caption{Generation of vector modes using a $q$-plate as described in the text (a) $q=+1/2$ plate (a1) $\mathrm{\bf{V}}_{+1,0}$ vector mode (a2) $\mathrm{\bf{V}}_{+1,\pi /2}$ vector mode. Also shown are the intensities of each vector mode as analyzed by a linear polarizer whose transmission axis is oriented at 0, 45, 90, and 135 degrees with respect to the lab table (a3) Schematic of the pattern of the liquid crystal molecules' orientations - black lines delineate orientation (a4) Higher-order Poincar\'e sphere representation for linear combination of $\mathrm{\bf{V}}_{+1,0}$ and $\mathrm{\bf{V}}_{+1,\pi /2}$ vector modes - double arrow indicates Ch1 and Ch2 as antipodal points (b) $q$=-1/2 plate (b1) $\mathrm{\bf{V}}_{-1,0}$ vector mode (b2) $\mathrm{\bf{V}}_{-1,\pi /2}$ vector mode (b3) Schematic of the pattern liquid crystal molecules' orientations (b4) Higher-order Poincar\'e sphere representation for linear combination of $\mathrm{\bf{V}}_{-1,0}$ and $\mathrm{\bf{V}}_{-1,\pi /2}$ vector modes - double arrow indicates Ch3 and Ch4 as antipodal points. Green arrows indicates light beam's direction of propagation.
}
\end{figure*}

Mode division multiplexing (MDM) is the method of optical communication where spatial modes are used as information channels carrying independent data streams. In general, MDM can potentially increase the information capacity of optical communication in an amount proportional to the number of modes used. MDM has been used in optical fiber communication; for comprehensive reviews see \cite{Richardson2013, Ryf2012}. Potentially, MDM can also be used in free space optical communication (FSO) \cite{Acampora2002, Willebrand2001}. Polarization division multiplexing and wavelength division multiplexing  have been used to increase the information capacity of FSO \cite{Cvijetic2010, Ciaramella2009}. 

	By definition, spatial modes are solutions to Maxwell's wave equation and can be represented by many bases \cite{Black2009}. In principle, any basis can be used for MDM. For example, what is sometimes referred to as the basis of orbital angular momentum (OAM) modes has been used to increase the information capacity of FSO and free space communication with millimeter waves \cite{Wang2012, Huang2014a, Yan}. There is also a basis sometimes referred to as the basis of vector modes or vector beams. Vector modes are spatial modes that have spatially inhomogeneous states of polarization, such as, radial and azimuthal polarization. They can produce smaller spot sizes and stronger longitudinal polarization components upon focusing. As a result, they are used for many applications, including optical trapping and nanoscale imaging \cite{Brown2, Zhan}.  
	
	In this work, vector modes are used to increase the information capacity of FSO via MDM. A mode (de)multiplexer for vector modes based on a liquid crystal technology referred to as a $q$-plate is introduced. As a proof of principle, using the mode (de)multiplexer four vector modes, each carrying a 20 Gbit/s quadrature phase shift keying signal on a single wavelength channel ($\lambda \sim $1550nm), comprising an aggregate 80 Gbit/s, were transmitted $\sim$1m over the lab table, with $<$-16.4 dB ($\sim 2 \%$) mode crosstalk. Bit error rates for all vector modes were measured at the forward error correction threshold with power penalties $<$ 3.41dB.

\section*{Vector modes}
				
	A vector mode is defined here as a solution to the wave equation in free space whose electric field is given by the equation \cite{Brown2,Zhan}:
\begin{eqnarray}
\mathrm{ \bf E }(r,\phi) &=& f(r) \mathrm{ \bf V}_{\ell, \gamma}(\phi),
\end{eqnarray}
where $(r, \phi )$ are cylindrical coordinates and $f(r)$ is a solution to the radial part of the wave equation, e.g., Bessel-Guassian \cite{Bessel}, or Laguerre-Gaussian functions. $\mathrm{ \bf  V}_{\ell, \gamma}(\phi)$ is a Jones vector describing the spatially inhomogeneous states of polarization of a vector mode given by the equation \cite{Stalder}:
\begin{eqnarray}
\mathrm{ \bf  V}_{\ell, \gamma}(\phi) &=& \left(\begin{array}{c} \cos( \ell \phi + \gamma ) \\ \sin( \ell \phi + \gamma ) \end{array}\right),
\end{eqnarray}
where $\ell = 0, \pm 1, \pm 2, ...$, and $\gamma = 0, \pi/2$. $\mathrm{ \bf  V}_{+1,0}$ and $\mathrm{ \bf  V}_{+1,\pi/2} $ are referred to as radial and azimuthal polarization and are shown in Fig. 1(a1) and Fig. 1(a2), respectively. $\mathrm{ \bf  V}_{-1,0}$ and $\mathrm{ \bf  V}_{-1,\pi/2}$ are shown in Fig. 1(b1) and Fig. 1(b2), respectively. The basis of vector modes comprises light's space and polarization degrees of freedom; the number of MDM channels that can be used for MDM when using vector modes is the same as using, for example, OAM modes together with polarization division multiplexing. 

\section{$q$-plate mode (de)multiplexer}

	To use vector modes for MDM, a mode (de)multiplexer for vector modes is required. In general, a mode (de)multiplexer (separates) combines $N$ spatial modes at the (receiver) transmitter of an optical communication system. For example, a mode (de)multiplexer for OAM modes that has been demonstrated is based on passive beam (separation) combining \cite{Wang2012, Huang2014a}. In passive beam (separation) combining, data streams from $N$ single mode optical fibers (SMFs) are transformed into $N$ OAM modes, or vice vera, via matching their wavefronts with $N$ spatial light modulators (SLMs). They are then (separated) combined via $N-1$ beam splitters. SLMs can also be used to generate vector modes \cite{Monika, Kimane}. However, there are many methods to generate vector modes; for example see \cite{Hasman1, Milionea, Milioneb, Milionec}. Here, a liquid crystal technology referred to as a $q$-plate is used \cite{Marruccia}. Analogous to how a SLM matches an OAM mode's wavefront, a $q$-plate matches a vector mode's spatially inhomogeneous state of polarization \cite{Marruccib}. A $q$-plate comprises a thin layer of pattern liquid crystal molecules in-between two thin glass plates. The molecules' orientations are described by $q \phi$, where $\phi$ is the azimuthal coordinate and $q$ is a half-integer. $q=+1/2$ and $q=-1/2$ plates are schematically shown in Fig. 1(a) and Fig. 1(b), respectively. Effectively, a $q$-plate is a half-wave plate with an azimuthally varying fast axis and can be mathematically represented by a Jones matrix \cite{Marruccia}:
\begin{equation}
\hat{Q} = \left(\begin{array}{cc} \cos( 2 q \phi )  & \sin( 2 q \phi) \\ \sin( 2 q \phi )  & -\cos( 2 q \phi ) \end{array}\right). 
\end{equation}
It can be shown using Jones calculus, upon propagation through a $q$-plate, the state of polarization of the fundamental mode of a SMF, given by the Jones vector $(\alpha \hspace{2mm} \beta)^T$ ($|\alpha|^2 + |\beta|^2 = 1; \alpha,\beta \in \mathbb{C}$, i.e., a linear combination of horizontal and vertical polarization) is transformed into a linear combination of vector modes, or vice versa, given by the equation:
\begin{eqnarray}
\alpha \mathrm{ \bf  V}_{2q, 0}(\phi) + \beta \mathrm{ \bf  V}_{2q, \pi/2}(\phi).
\end{eqnarray}
Using a $q=+1/2$ or $q=-1/2$ plate, horizontal/vertical polarization can be transformed into $\mathrm{ \bf  V}_{+1, 0}(\phi) / \mathrm{ \bf  V}_{+1, \pi/2}(\phi)$ or $\mathrm{ \bf  V}_{-1, 0}(\phi) / \mathrm{ \bf  V}_{-1, \pi/2}(\phi)$ vector modes, or vice versa, as shown in Fig. 1(a) and Fig. 1(b), respectively. 

Two $q=+1/2$ and $q=-1/2$ plates were fabricated. The fabrication details can be found in \cite{Marruccic, Marruccid}. Using the $q$-plates, a mode (de)multiplexer for vector modes based on passive beam (separation) combining was constructed as shown in Fig. 2. As a mode multiplexer- the fundamental modes of two SMFs at $\lambda \sim$1550nm were collimated to beam waists of $\sim$3mm using appropriate lenses (L) and then made to propagate through the $q$-plates. Then, the light beams from each SMF were aligned co-linearly via a non-polarizing beam splitter (BS). The $q$-plates were connected to an electronic signal generator generating a 1kHz square wave with a tunable voltage. The $q$-plates could be ``turned on" and ``turned off" by controlling the voltage. Each $q$-plate was ``turned on" ($\sim$1.2 Volts). Note, there is a -1.16dB loss when the $q$-plate is ``turned on." $\mathrm{ \bf  V}_{+1, 0}(\phi) / \mathrm{ \bf  V}_{+1, \pi/2}(\phi)$ and  $\mathrm{ \bf  V}_{-1, 0}(\phi) / \mathrm{ \bf  V}_{-1, \pi/2}(\phi)$ vector modes as generated using the mode (de)multiplexer are shown in Fig. 1(a1) /  Fig. 1(a2) and Fig. 1(b1) / Fig. 1(b2), respectively, along with their intensities as analyzed by a linear polarizer whose transmission axis is oriented at 0, 45, 90, and 135 degrees with respect to the lab.

\begin{figure}[htb]
\includegraphics[width=\linewidth]{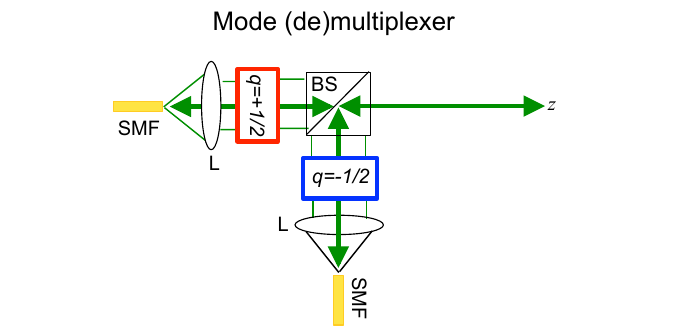}
\caption{Mode (de)multiplexer for vector modes based on passive beam (separation) combing as describe in text. Green arrows indicates the light beam's direction of propagation.
}
\end{figure}
\begin{figure*}[htb]
\includegraphics[width=\linewidth]{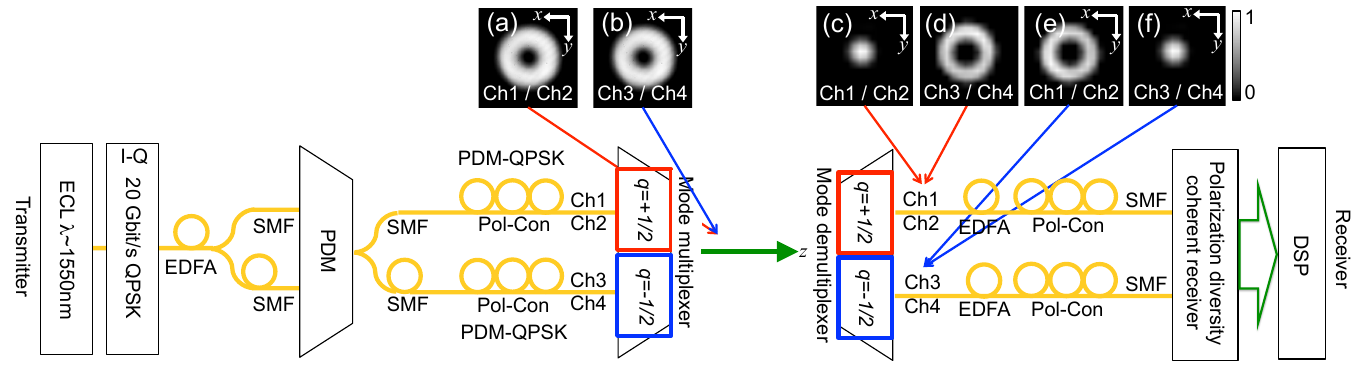}
\caption{Experimental setup as described in the text Intensity of (a) Ch1/Ch2 after $q=+1/2$ plate of mode multiplexer (b) Ch3/Ch4 after $q=-1/2$ plate of mode multiplexer (c) Ch1/Ch2 after $q=+1/2$ plate of mode demultiplexer (d) Ch3/Ch3 after $q=+1/2$ plate of mode demultiplexer (e) Ch1/Ch2 after $q=-1/2$ plate of mode demultiplexer (f) Ch1/Ch2 after $q=-1/2$ plate of mode demultiplexer. Green arrow indicates the light beam's direction of propagation.}
\end{figure*}
\section*{Mode division multiplexing using vector modes}

	Using the mode (de)multiplexer, four vector modes, each carrying a 20 Gbit/s quadrature phase shift keyed (QPSK) signal on a single wavelength channel ($\lambda \sim$1550nm), comprising an aggregate 80 Gbit/s, were transmitted as one light beam $\sim$1m over the lab table. Note, over this distance there is negligible atmospheric turbulence and beam divergence. The experimental setup is shown in Fig. 2. The transmitter was as follows. First, the output of an external cavity laser (ECL) tuned to $\lambda\sim1550$nm was modulated by an I-Q modulator to generate a 20 Gbit/s QPSK signal. The I-Q modulator was driven by a two-channel pattern generator generating two 10 Gbaud signals, being decorrelated pseudo-random-binary-bit-sequences of length $2^{15} - 1$, and comprising the in-phase (I) and quadrature (Q) components of the QPSK signal. The signal was then amplified by a low-noise erbium doped fiber amplifier (EDFA) and polarization division multiplexed (PDM) by splitting and decorrelating it in two SMFs, making the states of polarization in each SMF mutually orthogonal, (horizontal/vertical polarization), and then recombining them into one SMF. Finally, the resultant PDM-QPSK signal was mode multiplexed by again splitting and decorrelating it in two SMFs which were then connected to the SMFs of the mode (de)multiplexer, resulting in four vector modes each carrying a 20 Gbit/s QPSK signal.
		
	At the mode (de)multiplexer, due to twists and bends in the SMFs, the channels of each PDM-QPSK signal are not necessarily horizontal and vertical polarization. However they can be aligned using a polarization paddle controller (Pol-Con). Nonetheless, each channel is a linear combination of horizontal and vertical polarization as represented by two arbitrary antipodal points on the Poincar\`e sphere. Therefore, as given by Eq. 5, after propagation through a $q$-plate each channel of a PDM-QPSK signal is transformed into a linear combination of vector modes as represented by two arbitrary antipodal points on a higher-order Poincar\`e sphere \cite{Milione2011, Milione2, Aiello}. The channels originating from the $q=+1/2$ and $q=-1/2$ plates are labelled Ch1/Ch2 and Ch3/Ch4 as shown in Fig. 1(a4) and Fig. 1(b4), respectively. The intensities of Ch1/Ch2 and Ch3/Ch4, as recorded using the InGaAs camera, are shown in Fig. 3(a) and Fig. 3(b), respectively. 

	After transmission over the lab table, the receiver was as follows. First, the channels were mode demultiplexed into the two SMFs of another mode (de)multiplexer. Note, Ch1/Ch2 and Ch3/Ch4 were received and discriminated via coupling to both SMFs. In accordance with Eq. 3, upon propagation through a $q=+1/2$ plate, $\mathrm{ \bf  V}_{+1, 0}(\phi) / \mathrm{ \bf  V}_{+1, \pi/2}(\phi)$ vector modes are transformed into horizontal/vertical polarization while $\mathrm{ \bf  V}_{-1, 0}(\phi) / \mathrm{ \bf  V}_{-1, \pi/2}(\phi)$ vector modes are transformed into higher-order vector modes. The intensities of Ch1/Ch2 and Ch3/Ch4 at the SMF corresponding to the $q=+1/2$ $q$-plate, as recorded using an InGaAs camera, are shown in Fig. 3(c) and Fig. 3(d), respectively. As can be seen, Ch1/Ch2 were transformed back into the fundamental mode of a SMF, i.e, a PDM-QPSK signal, and were coupled to the SMF. However, Ch3/Ch4 were transformed into higher-order vector modes and were not coupled to the SMF. The opposite is true for the $q=-1/2$ $q$-plate. The intensities of Ch1/Ch2 and Ch3/Ch4 at the SMF corresponding to the $q=-1/2$ $q$-plate, as recorded using an InGaAs camera, are shown in Fig. 3(e) and Fig. 3(f), respectively. Next, each resulting PDM-QPSK signal was amplified by an EDFA and aligned via a Pol-Con to a polarization diversity coherent receiver where they were polarization demultiplexed and coherently detected. The polarization diversity coherent receiver comprised a polarizing splitter, followed by two optical hybrids, connected to another ECL ("local oscillator"). Via intradyne detection, the analog QPSK signals were converted to digital signals by four balanced detectors. Using four digital oscilloscopes, the resultant digital signals were captured at 40 GSample/s ($\Delta$20 GHz) for offline digital signal processing (DSP). The captured constellations of the QPSK signals for all channels are shown in Fig 3(a). 	

\section*{Mode crosstalk, Bit error rates, and power penalties}
	
	Mode crosstalk (MC) is defined as the transfer of power (signals) between channels. MC for each channel was measured at the mode demultiplexer by transmitting one channel at a time and measuring the power received by each of all channels. A normalized MC ``matrix", comprising the MC measurements, are shown in Fig. 3(b). There is $<$-16.7dB ($\sim2\%$) MC for any channel. As can be seen, via the off-diagonal blocks, the greatest contribution to MC for any channel came from channels originating from different $q$-plates. As there is negligible atmospheric turbulence and beam divergence over the transmitted distance, this is attributed to mismatch of the numerical apertures and misalignment of the lenses with the SMFs of the mode demultiplexer. MC may also originate from ``mode impurity," i.e., power is generated in vector modes other than those intended. A preliminary analysis indicates the vector modes as generated via the $q$-plates of the mode multiplexer have $< - 21.0$dB $(<1\%)$ mode impurity. A detailed analysis of the measurement of mode impurity will be reported elsewhere.

	For each channel and a``back to back" (B2B) channel, bit error rates (BER) were measured as a function of optical signal to noise ratio (OSNR) when all channels were simultaneously transmitted. BER vs. OSNR curves are shown in Fig. 3(a). The B2B channel is the original QPSK signal, transmitted and received, when the $q$-plates are ``turned off." The performance of each channel can be assessed via its power penalty (PP), i.e., the difference in OSNR between each channel and the B2B channel at the forward error correction (FEC) threshold (BER = $3.8\times10^{-3}$). The PPs for all channels and their cumulative MCs are shown in Table 1. All channels reach a BER at the FEC threshold with a PP $<$ 3.41dB. Differences in PPs are attributed to polarization dependent loss throughout the various SMFs of the system and misalignment of the Pol-Cons. 

\section*{Conclusion}

	In conclusion, vector modes were used to increase the information capacity of FSO via MDM. A mode (de)multiplexer for vector modes based on a liquid crystal technology referred to as a $q$-plate was introduced. As a proof of principle, using the mode (de)multiplexer four vector modes, each carrying a 20 Gbit/s quadrature phase shift keying signal on a single wavelength channel ($\lambda \sim $1550nm), comprising an aggregate 80 Gbit/s, were transmitted $\sim$1m over the lab table, with $<$-16.4 dB ($\sim 2 \%$) mode crosstalk. BERs for all vector modes were measured at the forward error correction threshold with power penalties $<$ 3.41dB.	
	
	It has been conjectured upon propagation through atmospheric turbulence a vector mode will experience less scintillation as compared to an OAM mode \cite{Haus, Gbur}. The propagation of vector modes through atmospheric turbulence is the subject of future work. When using OAM modes for MDM, the deleterious effects of atmospheric turbulence can be compensated via wavefront measurements \cite{Ren}. When using vector modes for MDM, comparable compensation of atmospheric turbulence may be possible via the measurement of spatially inhomogeneous polarization \cite{Angela}.

	It has been shown, the transmission of an OAM mode through scattering media depends on its polarization and OAM \cite{Milione4}. As vector modes are non-separable superpositions of circular polarized OAM modes \cite{Milione2011,Milione2, Aiello, spinorbit}, a comparable study of vector modes may be of interest, especially with regard to atmospheric light scattering. 
	
	Vector modes are referred to as the ``true modes" of an optical fiber. Use of the $q$-plate mode (de)multiplexer for optical fiber communication, as well as its compatibility with wavelength division multiplexing, especially with regard to ``ring core" optical fibers, is the subject of current work \cite{Milione1, Milione5, Milione6, Nenad, Leslie}. 

	It is noted, there are other light beams that have spatially inhomogeneous states of polarization, such as, Full Poincare beams \cite{Beckley}. However, as compared to vector modes, Full Poincare beams experience non-trivial dynamics upon propagation \cite{Milione3,Hasman2}. 	
\begin{figure}[htb]
\includegraphics[width=\linewidth]{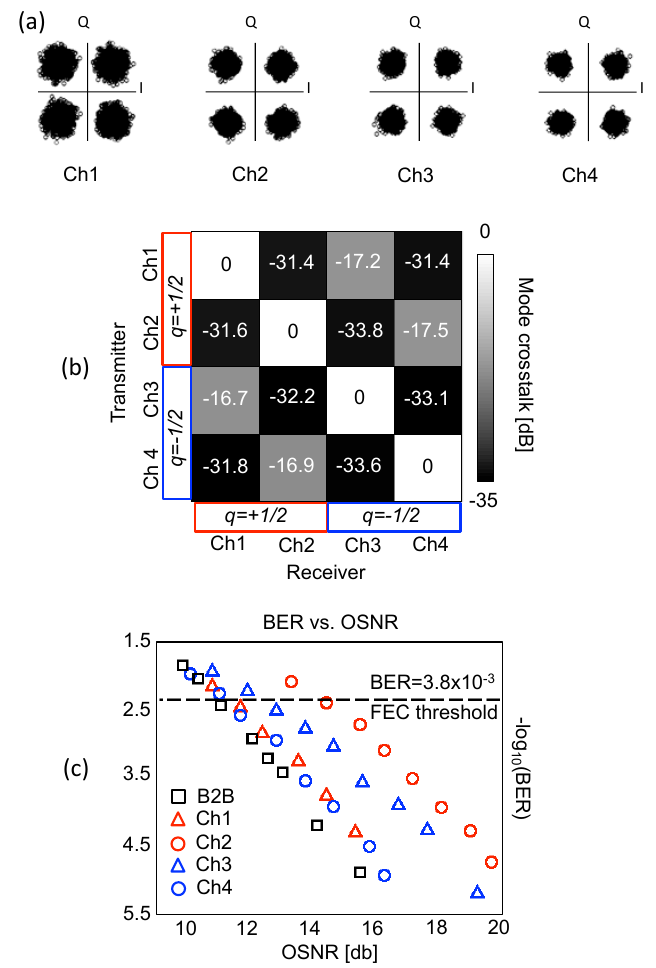}
\caption{(a) Quadrature phase shift keying (QPSK) constellations for Ch1, Ch2, Ch3, and Ch4 (b) Mode crosstalk matrix for Ch1, Ch2, Ch3, and Ch4 (c) Measurements of bit error rate (BER) as a function of optical signal to noise ratio (OSNR) for Ch1, Ch2, Ch3, Ch4, and a back to back (B2B) channel. Dashed line delineates forward error correction (FEC) threshold.}
\end{figure}
\begin{table}[htbp]
\centering
\caption{\bf Power Penalties}
\begin{tabular}{ccc}
\hline
Vector mode & Power Penalty [dB] & Mode Crosstalk [dB]  \\
\hline
Ch1 & 0.44 & -16.42 \\
Ch2 & 3.40 & -16.64 \\
Ch3 & 1.46 & -17.01 \\
Ch4 & 0.41 & -17.21 \\
\hline
\end{tabular}
\end{table}

\section*{Funding Information}
GM and RRA acknowledge support from AFOSR Grant. No.  No. 47221-00-01, ARO Grant. No. 52759-PH-H, NSF GRFP Grant. No. 40017-00-04, and Corning, Inc. EK acknowledges support from the Canada Excellence Research Chairs (CERC) Program. MPJL acknowledges support from EPSRC and the Royal Academy of Engineering. USC acknowledges support from the DARPA InPho program.

\section*{Acknowledgments}
GM thanka S. Slussarenko for fabricating the $q$-plates. Authors from USC thank Dr. Prem Kumar and Dr. Tommy Willis for helpful discussions.

\end{document}